\documentclass[letterpaper]{article}
\usepackage{aaai}
\usepackage{times}
\usepackage{helvet}
\usepackage{courier}
\usepackage{graphicx}
\usepackage{epstopdf}
\usepackage{amsfonts}
\usepackage{amsmath}
\usepackage{multirow}
\usepackage{bm}
\usepackage{color}

\usepackage{hyperref}
\usepackage[backend=biber,style=apa]{biblatex}
\DeclareLanguageMapping{english}{english-apa}
\begin{document}
%
\title{HIP: Model-Agnostic Hypergraph Influence Prediction via Distance-Centrality Fusion and Neural ODEs}

\author{Su-Su Zhang$^{\rm 1}$, JinFeng Xie$^{\rm 1}$,  Yang Chen$^{\rm 2}$, Min Gao$^{\rm 2}$, Cong Li$^{\rm 3}$, Chuang Liu$^{\rm 1 *}$, Xiu-Xiu Zhan$^{\rm 1 *}$\\
\textsuperscript{\rm 1}Research Center for Complexity Sciences, Hangzhou Normal University, Hangzhou 311121, PR China \\
\textsuperscript{\rm 2}Shanghai Key Lab of Intelligent Information Processing,
School of Computer Science, Fudan University, Shanghai 200433, PR China\\
\textsuperscript{\rm 3}Adaptive Networks and Control Laboratory, Electronic Engineering Department, School of Information Science and Engineering, \\ and the Research Center of Smart Networks and Systems, Fudan University, Shanghai 200433, PR China\\
$^{\rm *}$ Corresponding authors. Email address: liuchuang@hznu.edu.cn (Chuang Liu), zhanxiuxiu@hznu.edu.cn (Xiu-Xiu Zhan)
}

\maketitle
\begin{abstract}
\begin{quote}

Predicting user influence in social networks is a critical problem, and hypergraphs, as a prevalent higher-order modeling approach, provide new perspectives for this task. However, the absence of explicit cascade or infection probability data makes it particularly challenging to infer influence in hypergraphs. To address this, we introduce HIP, a unified and model-independent framework for influence prediction without knowing the underlying spreading model. HIP fuses multi-dimensional centrality indicators with a temporally reinterpreted distance matrix to effectively represent node-level diffusion capacity in the absence of observable spreading. These representations are further processed through a multi-hop Hypergraph Neural Network (HNN) to capture complex higher-order structural dependencies, while temporal correlations are modeled using a hybrid module that combines Long Short-Term Memory (LSTM) networks and Neural Ordinary Differential Equations (Neural ODEs). Notably, HIP is inherently modular: substituting the standard HGNN with the advanced DPHGNN, and the LSTM with xLSTM, yields similarly strong performance, showcasing its architectural generality and robustness. Empirical evaluations across 14 real-world hypergraph datasets demonstrate that HIP consistently surpasses existing baselines in prediction accuracy, resilience, and identification of top influencers, all without relying on any diffusion trajectories or prior knowledge of the spreading model. These findings underline HIP's effectiveness and adaptability as a general-purpose solution for influence prediction in complex hypergraph environments.

\end{quote}
\end{abstract}

\section{Introduction}

The objective of influence prediction is to estimate the extent to which a user's message can propagate through a social network, based on user-specific features and structural connectivity~\parencite{shang2025leveraging}. Accurate prediction of node influence plays a vital role in a range of downstream tasks, such as increasing product exposure through information diffusion to drive commercial gains~\parencite{yang2024comparative}, enhancing the precision of rumor suppression techniques~\parencite{bian2020rumor}, and enabling early interventions in epidemic control by uncovering super-spreaders hidden within group-level contact structures~\parencite{morris2021optimal}.

Existing influence prediction methods in conventional pairwise networks can be broadly categorized into four types: Monte Carlo (MC) simulations, sampling-based approaches~\parencite{cheng2013staticgreedy, ohsaka2014fast, borgs2014maximizing}, probabilistic inference models~\parencite{jiang2011simulated, aghaee2020efficient}, and neural network-based algorithms~\parencite{cheng2024information, aravamudan2023anytime}. While MC simulations offer high accuracy, they are computationally expensive and impractical for large-scale networks. To improve efficiency, sampling-based methods such as snapshot-based~\parencite{lu2015influence, lu2016big} and reverse influence sampling~\parencite{nguyen2017importance, tang2018online} have been proposed, though they face a trade-off between accuracy and scalability. Probabilistic inference methods, inspired by the two-hop influence theory~\parencite{pei2014searching}, perform well in sparse networks but often lose accuracy in denser settings~\parencite{gong2016influence}. To address structural complexity and scalability, neural network-based models have emerged~\parencite{qiu2018deepinf}, though extracting meaningful node-level influence features remains challenging. Recent advances include centrality-based~\parencite{kumar2023influence}, random walk-based~\parencite{wang2024adaptive}, diffusion-based~\parencite{panagopoulos2021}, and infection probability-based embeddings~\parencite{yuan2024graph}.

Despite advances in influence prediction on conventional networks, two key limitations persist. First, most models ignore \textbf{higher-order structures}, which are common in real-world systems~\parencite{battiston2020networks}. Hypergraphs naturally capture these multi-node interactions that significantly alter diffusion dynamics~\parencite{ferraz2024contagion}, yet remain underexplored in existing models. Second, \textbf{spreading trajectories and infection probabilities} are often unavailable in practice. For example, during the COVID-19 pandemic, contact tracing was difficult despite case data being accessible. Methods relying on such detailed information~\parencite{yuan2024graph} face limited applicability. These challenges call for predictive models that operate solely on structural data without requiring full diffusion traces.

To address the aforementioned challenges, we propose HIP (\textbf{H}ypergraph \textbf{I}nfluence \textbf{P}rediction), a novel general framework designed for influence prediction without knowing the underlying spreading model in hypergraphs. HIP comprises four key components: (1) Distance-Centrality Feature Fusion, (2) Contextual Encoding via Multi-hop Hypergraph Propagation, (3) Spatiotemporal Dynamics Modeling via LSTM-ODE Integration, and (4) Diffusion Magnitude Prediction. In the first module, we reinterpret the node distance matrix from a temporal perspective by treating each entry as the minimal number of hops required for information spread, thereby assigning temporal semantics to structural distances. Additionally, we compute multi-dimensional centrality metrics to capture local to global structural information based on higher-order interactions. In the second module, we leverage the classic HGNN~\parencite{feng2019hypergraph} to aggregate features from a node's two-hop neighborhood by modeling message passing between nodes and hyperedges. Next, to capture dynamic propagation patterns, we integrate two temporal models: LSTM ~\parencite{hochreiter1997long} and Neural ODEs~\parencite{chen2018neural}, which further refine the time-aware representations. The current implementation of HIP adopts HGNN and LSTM as its default components, which can be seamlessly substituted with more advanced architectures. Finally, a Multilayer Perceptron is used in the Diffusion Magnitude Prediction module to estimate each node's final influence. We evaluate HIP on 14 empirical hypergraphs against a suite of SOTA baselines. Experimental results show that HIP consistently achieves superior performance in terms of prediction accuracy, influential node identification, and robustness. The main contributions of this work are summarized as follows.
\begin{itemize}

   \item We formulate the novel problem of influence prediction on hypergraphs, which poses significant challenges for conventional methods due to the presence of complex and unobservable higher-order spreading models.

   \item We develop an end-to-end prediction framework (HIP) based on HNNs and Neural ODEs to predict node influence, which effectively captures the higher-order structural dependencies and the temporal dynamics of influence propagation. 

  \item Extensive experiments on real-world and synthetic hypergraphs show that HIP consistently outperforms state-of-the-art baselines in accuracy, robustness, and influential node identification, without requiring explicit diffusion data.
\end{itemize}


    

\section{Preliminaries}

\textbf{Definition 1 (Hypergraph)} A hypergraph is defined as $H = (V, E)$, where $V = \{v_1, v_2, \cdots, v_N\}$ denotes the set of nodes, and $E = \{e_1, e_2, \cdots, e_M\}$ represents the set of hyperedges. Each hyperedge corresponds to a group-based interaction involving multiple nodes. To characterize the associations between nodes and hyperedges, we construct an incidence matrix $\mathbf{I} \in \mathbb{R}^{N \times M}$, where $\mathbf{I}_{ij} = 1$ if user $v_i$ participates in hyperedge $e_j$, and $\mathbf{I}_{ij} = 0$ otherwise. Based on $\mathbf{I}$, we derive the corresponding adjacency matrix $\mathbf{A} \in \mathbb{R}^{N \times N}$, where $\mathbf{A}_{ij} = 1$ indicates that users $v_i$ and $v_j$ co-occur in at least one hyperedge.

\textbf{Definition 2 (Influence Prediction)} When a user $v_i$ initiates a post in a social network, the information propagates implicitly through the network structure, resulting in a subset of users eventually receiving and endorsing the content. When the social network is modeled as a hypergraph, the final influence of user $v_i$ is measured by the total number of users who have endorsed the post, denoted as $I_i$. Given a hypergraph $H$ and an initial user $v_i$, the objective is to develop a predictive function $f(H, v_i)$ that estimates the final influence value $I_i$,
\begin{equation}
f(H, v_i) \longrightarrow I_i.
\end{equation}

\section{Design and Implementation of HIP framework}
\begin{figure*}
    \centering
    \includegraphics[width=0.8\linewidth]{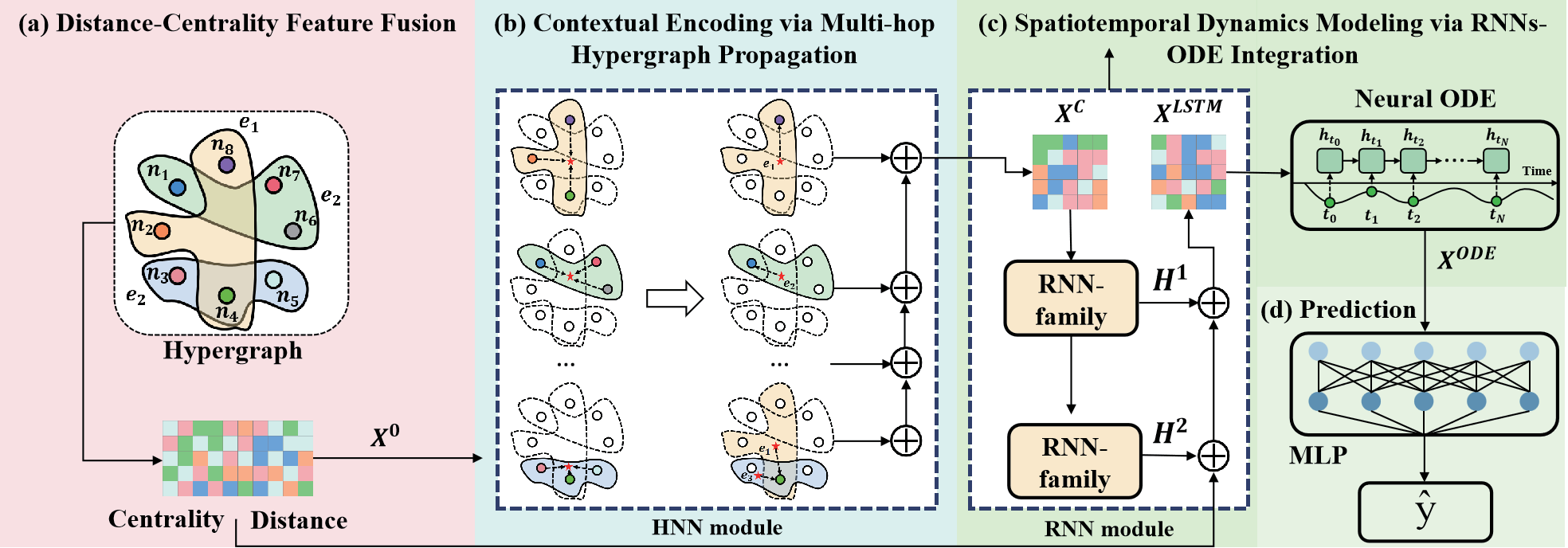}
    \caption{\textbf{Architectural overview of the HIP model.} \textbf{(a)} Encodes node propagation capacity via multi-dimensional centrality and distance features. \textbf{(b)} Aggregates neighborhood information through two-hop HNNs module propagation. \textbf{(c)} Learns latent temporal patterns from higher-order distances via RNN module to enhance prediction. \textbf{(d)} Outputs the predicted influence value for each node.}
    \label{fig:HGE_framework}
\end{figure*}

In this section, we provide a detailed description of the proposed method, i.e., HIP, whose overall architecture is depicted in Figure~\ref{fig:HGE_framework}. The HIP framework consists of four main components, described as follows:

\begin{itemize}
    \item \textbf{Distance-Centrality Feature Fusion} is designed to capture informative characteristics that reflect the propagation capacity of individual nodes. It integrates centrality and distance-based features from both local and global perspectives to construct a comprehensive multi-level representation.
    
    \item \textbf{Contextual Encoding via Multi-hop Hypergraph Propagation} enables each node to exchange initial features with its one-hop and two-hop neighbors through HNNs, thereby capturing rich contextual dependencies. 

    \item \textbf{Spatiotemporal Dynamics Modeling via RNNs-ODE Integration} leverages the higher-order distance matrix to capture latent temporal dynamics, thereby improving the accuracy of influence prediction.

    \item \textbf{Diffusion Magnitude Prediction} returns the predicted final influence of each node based on a Multilayer Perceptron. 
    
\end{itemize}

\subsection{Distance-Centrality Feature Fusion}
We integrate node-level distance with hypergraph-based centrality measures to construct the initial feature matrix. Specifically, the distance matrix $\mathbf{D}$ is derived from the shortest path lengths between nodes in the underlying conventional network of the hypergraph. Notably, this distance matrix exhibits temporal characteristics: each row vector $\mathbf{D}_i = (d_v(i,1), d_v(i,2), \dots, d_v(i,N))$ can be interpreted as the minimum time required for information originating from node $v_i$ to reach all other nodes. This perspective offers a novel interpretation of distance as a proxy for information diffusion latency, thus bridging topological structure and spreading dynamics. The Appendix outlines the hypergraph centrality measures employed in our framework.

According to the distance matrix and multi-dimensional centrality, we obtain the initial feature matrix $\mathbf{X}^{(0)} \in \mathbb{R}^{N \times (N+5)}$, in which the feature vector of node $v_i$ is given as $\mathbf{X}^{(0)}_i = (d_v(i,1), d_v(i,2), \dots, d_v(i,N), k_i, k_i^H, c_i, g_i, h_i)$.

\subsection{Contextual Encoding via Multi-hop Hypergraph Propagation}
Inspired by the observation of Pei et al.~\parencite{pei2014searching}, which indicates that a node's influence predominantly resides within its two-hop neighborhood, we adopt a two-layer HGNN to facilitate localized feature aggregation. Each HGNN layer operates under a message-passing framework and consists of two sequential transformation steps designed to effectively encode structural context. In the first step, each hyperedge gathers features from its associated nodes to construct the hyperedge-level representation $\mathbf{Z}^{(l)}_e$
\begin{equation}
    \mathbf{Z}^{(l)}_e= \mathbf{W}_e \mathbf{(K^E)}^{-1} \mathbf{I}^\top \mathbf{(K^H)}^{-\frac{1}{2}} \mathbf{X}^{(l)},
\end{equation}
where $\mathbf{W}_e$ is the diagonal matrix of hyperedge weights, typically initialized as the identity matrix. $\mathbf{(K^E)}^{-1}$ and $\mathbf{(K^H)}^{-\frac{1}{2}}$ correspond to the symmetrically normalized hyperedge size and hyperdegree matrices, respectively. $\mathbf{X}^{(l)}$ represents the node feature matrix at the $l$-th layer. Next, each node updates its representation by aggregating information from all hyperedges it participates in. The corresponding formulation is given by
\begin{equation}
    \mathbf{X}^{(l+1)} = \sigma \left( \mathbf{(K^H)}^{-\frac{1}{2}} \mathbf{I} \mathbf{W}_e \mathbf{(K^E)}^{-1}
        \mathbf{I}^\top \mathbf{(K^H)}^{-\frac{1}{2}} \mathbf{X}^{(l)} \mathbf{\Theta}^{(l)} \right),
\end{equation}
where $\mathbf{X}^{(l+1)}\in \mathbb{R}^{N \times d}$ denotes the updated node feature matrix and $\sigma$ denotes the nonlinear activation function.  The learnable weight matrix $\mathbf{\Theta}^{(l)} \in \mathbb{R}^{C \times d}$ maps input features of dimension $C$ to output features of dimension $d$. After passing through the first and second HGNN layers, the resulting outputs $\mathbf{X}^1$ and $\mathbf{X}^2$ are concatenated to form the final node representation $\mathbf{X}^c = \text{Concat}(\mathbf{X}^1, \mathbf{X}^2)$.

\subsection{Spatiotemporal Dynamics Modeling via RNNs-ODE Integration}
To capture the temporal dynamics embedded in relative distance patterns, thereby modeling how information potentially propagates over time, we integrate LSTM networks with Neural ODEs to process the feature matrix $\mathbf{X}^c \in \mathbb{R}^{N\times 2*d}$. In the first stage, $\mathbf{X}^c$ is passed through a two-layer LSTM to learn sequential dependencies inherent in the distance-based representations
\begin{equation}
\left[ \mathbf{X}^{LSTM},\ \mathbf{H}^{1}, \mathbf{H}^{2} \right] = \text{LSTMs}(\mathbf{X^c}).
\end{equation}
The output from the first LSTM layer, denoted as $\mathbf{X}^{L1}$, is subsequently fed into a second LSTM layer. We expect the hidden states $\mathbf{H}^{1}$ and $\mathbf{H}^{2}$ from the two LSTM layers to effectively capture the latent diffusion dynamics, as encoded through distance-based representations. To preserve the original structural information and enrich the expressive power of the model, we concatenate the initial node features $\mathbf{X}^0$ with the LSTM outputs, resulting in the fused representation $\mathbf{X}^{LSTM} = \text{Concat}(\mathbf{H}^{1}, \mathbf{H}^{2}, \mathbf{X}^0)$.


To capture continuous-time dynamics in hypergraphs, we employ the Neural ODE module that replaces the discrete-layer architectures with differential equations. Taking $\mathbf{X}^{LSTM}$ as the input, the evolution of the hidden state over continuous time $t \in [t_0, t_1]$ is governed by the following ordinary differential equation $\frac{d\mathbf{X}(t)}{dt}=f(\mathbf{X}(t), t; \theta), \mathbf{X}(t_0)=\mathbf{X}^{LSTM}$,
where $f(\cdot)$ is a neural network parameterized by $\theta$. The final state after integration, denoted as $\mathbf{X}^{ODE}$, is obtained by solving the following ODE $\mathbf{X}^{ODE}=odeint(f, \mathbf{X}^{LSTM}, t_0, t_1)$,
where $odeint$ is the ODE solver provided by the TorchDiffEq library~\parencite{torchdiffeq}. We fix the integration interval to $[t_0, t_1] = [0, 1]$, enabling the model to learn temporal transitions over a standardized time window, which simplifies the numerical computation and ensures consistency across different input samples.

\subsection{Diffusion Magnitude Prediction}
The final influence prediction $\hat{y}$ is derived by applying two Multilayer Perceptrons, each followed by ReLU activation and dropout, to the learned node representation $\mathbf{X}^{ODE}$
\begin{equation}
    \hat{y}=\sigma(\mathbf{W}_2 (\sigma(\mathbf{W}_1 \mathbf{X}^{ODE}+b_1))+b_2),
\end{equation}
where $\sigma(\cdot)$ denotes the ReLU activation, $\mathbf{W}$ and $b$ represent the weights and bias of the Multilayer Perceptron., respectively.

To quantify the discrepancy between the predicted value $\hat{y_i}$ and the ground truth $y_i$ for each node $v_i$, we employ the Mean Squared Logarithmic Error (MSLE) as the loss function $\mathcal{L}$, which guides the optimization of model parameters. The formulation of $\mathcal{L}$ is given by
\begin{equation}
\label{equ:MSLE}
\mathcal{L} = \frac{1}{N} \sum_{i=1}^N \left( \log(\hat{y}_i + 1) - \log(y_i + 1) \right)^2.
\end{equation}

\section{Experiments}
In this section, we conduct a thorough evaluation of the proposed HIP algorithm in comparison with SOTA approaches across multiple dimensions. Specifically, we aim to address the following key questions:
\textbf{(1)} How well does our method perform in predicting node influence, particularly under scenarios where the underlying diffusion model is unknown?
\textbf{(2)} To what extent do distance-based and centrality-driven features contribute to the accuracy of influence prediction?
\textbf{(3)} Can the HIP framework be effectively applied to downstream tasks, such as identifying influential nodes within a hypergraph?

A portion of the experimental results has been moved to the Appendix for clarity and space considerations, including the Tables~\ref{tab:kend_ICRP_threshold} and \ref{tab:MSLE_threshold}.

\begin{table}[t]
\caption{\label{topological properties} \small \textbf{Topology properties of empirical hypergraphs.} We show the the number of nodes $N$, number of hyperedges $M$, average node degree $\langle k \rangle$, average node hyperdegree $\langle k^H \rangle$, average hyperedge size $\langle k^E \rangle$ and coefficient of variation of node degree distribution $CV(\langle k \rangle)$.}
	\centering
	\label{tab:topological properties} 
    {\fontsize{7pt}{8pt}\selectfont
    \setlength{\tabcolsep}{3pt}
    {
	\begin{tabular}{@{}ccccccc@{}}
		\hline\hline\noalign{\smallskip}	
         \textbf{Hypergraphs} & \textbf{$N$} & \textbf{$M$} & \textbf{$\langle k \rangle$} & \textbf{$\langle k^H \rangle$} & \textbf{$\langle k^E \rangle$} & $CV(\langle k \rangle)$ \\
        \noalign{\smallskip}\hline\noalign{\smallskip}
        \textbf{Alg}~\parencite{amburg2020fair} & 423 & 1268 & 78.9 & 19.53 & 6.52 & 0.87 \\
         \textbf{Geo}~\parencite{amburg2020fair} & 580 & 1193 & 164.79 & 21.53 & 10.47 & 0.74 \\
         \textbf{Bars}~\parencite{amburg2020fair} & 1234 & 1194 & 174.3 & 9.62 & 9.94 & 0.83 \\
         \textbf{Res}t~\parencite{amburg2020fair} & 565 & 601 & 79.75 & 8.14 & 7.66 & 0.75 \\
         \textbf{Music}~\parencite{ni2019justifying} & 1106 & 694 & 167.88 & 9.49 & 15.13 & 0.64 \\
         \textbf{iAF}~\parencite{yadati2020nhp} & 1668 & 2351 & 13.26 & 5.46 & 3.87 & 3.30 \\
         \textbf{iJO}~\parencite{yadati2020nhp} & 1805 & 2546 & 16.92 & 5.55 & 3.94 & 3.07 \\
         \textbf{SenateCm}~\parencite{chodrow2021generative} & 282 & 315 & 100.77 & 19.18 & 17.17 & 0.44 \\
         \textbf{HouseCm}~\parencite{chodrow2021generative} & 1290 & 341 & 195.56 & 9.18 & 34.73 & 0.54 \\
         \textbf{HS2012}~\parencite{mastrandrea2015contact} & 180 & 2911 & 24.67 & 40.81 & 2.52 & 0.44 \\
         \textbf{HS2013}~\parencite{mastrandrea2015contact} & 327 & 8644 & 35.58 & 79.43 & 3.00 & 0.38 \\
         \textbf{ht09}~\parencite{isella2011s} & 113 & 2539 & 38.87 & 58.82 & 2.62 & 0.47 \\
         \textbf{SocHam}~\parencite{nr} & 2426 & 5041 & 13.71 & 10.63 & 5.11 & 1.45 \\
         \textbf{FBTV}~\parencite{nr} & 3892 & 6355 & 8.86 & 7.23 & 4.43 & 1.42 \\
        \hline\hline\noalign{\smallskip}
	\end{tabular}
    }
    }
\end{table}

\subsection{Datasets}
We evaluate the proposed method on fourteen real-world hypergraph datasets exhibiting diverse structural characteristics, as detailed in Table~\ref{tab:topological properties}. These datasets span a variety of domains and are frequently used in classical learning tasks, such as link prediction, community detection and node clustering, ensuring the generality of our analysis.  For each dataset, we randomly partition the data into training, validation, and test sets following a $70\%:20\%:10\%$ split.

\subsection{Baselines}
The baseline methods that are used for comparison could be classified into three categories: \textbf{(1) Existing Prediction Methods}: To assess accuracy, we compare with methods designed to predict node influence spread, including \textbf{MPNN+LSTM}~\parencite{panagopoulos2021} employs the spreading sequence over time steps as additional information; \textbf{DeepIM}~\parencite{ling2023deep} is a two-stage model, dividing into spreading probability prediction and seed node selection. \textbf{GBIM}~\parencite{yuan2024graph} models diffusion via a scalable surrogate integrating kernel attention and Bayesian regression; and \textbf{ALGE}~\parencite{zhu2023precise} estimates the influence of nodes in the social networks by an active learning architecture. 
\textbf{(2) Temporal Neural Networks}: To evaluate the effectiveness of temporal modeling, we include \textbf{GLSTM}~\parencite{kumar2023influence} and \textbf{Bi-GRU}~\parencite{cho2014learning}, which are standard sequence models for capturing temporal dependencies.
\textbf{(3) Efficient Centrality Measures}: To test ranking consistency, we compare with structural metrics \textbf{HCI}~\parencite{morone2015influence} and \textbf{H-index}~\parencite{lu2016h}, which quantify node influence based on local topology.


\subsection{Evaluation metrics}
To quantitatively evaluate model performance, we adopt Kendall's $\tau$ correlation coefficient, Top-$K$ Overlap ($O$), Log-Transformed $R^2$ (Log-$R^2$), Mean Squared Logarithmic Error (MSLE) and Mean Relative Logarithmic Error (MRLE) as our primary evaluation metrics. These evaluation metrics are chosen due to their widespread use in influential studies on influence prediction, such as~\parencite{esfandiari2024predicting, kumar2022influence}


\textbf{Kendall's $\tau$}~\parencite{kendall1938new} quantifies the rank correlation between two orderings. A higher $\tau (\tau \in [-1,1])$ value indicates better alignment between predicted and true rankings.

\textbf{Top-$K$ Overlap ($O$)} measures the consistency between predicted and ground-truth top-$K$ node sets, i.e., $O = \frac{|S \cap \hat{S}|}{N}$, where $S$ and $\hat{S}$ represent the sets of top-$K$ nodes identified by the ground truth and the prediction, respectively.

\textbf{Log-$R^2$} assesses the fit between $\log(1 + y_i)$ and $\log(1 + \hat{y}_i)$, i.e., $R^2_{\log} = 1 - \frac{\sum_i \left( \log(1 + y_i) - \log(1 + \hat{y}_i) \right)^2}{\sum_i \left( \log(1 + y_i) - \overline{\log(1 + y)} \right)^2}$.

\textbf{MSLE} computes the mean squared difference of log-transformed predictions and labels (see Eq.~\ref{equ:MSLE}).

\textbf{MRLE} quantifies the mean relative log error $\text{MRLE} = \frac{1}{n} \sum_i \left| \frac{\log(1 + \hat{y}_i) - \log(1 + y_i)}{\log(1 + y_i)} \right|$.

\subsection{Hypergraph-based spreading models}
To better model higher-order interactions inherent in real-world diffusion processes, we extend the classic Independent Cascade (IC) model to hypergraphs, resulting in the IC model with Reactive Process and threshold Constraint (ICRP-$\lambda$). In this model, nodes are either active or inactive. For each hyperedge $e_m$, if the fraction of active nodes reaches or exceeds a threshold $\lambda$, the active nodes can activate inactive nodes in adjacent hyperedges at the next step with probability $p$. Once activated, a node remains active permanently but can activate its neighbors only once.  The specific spreading process is as follows: The diffusion process starts with an initial seed node $v_i$ set to the active state. Let $\mathcal{S}_{t-1}$ denote the set of nodes newly activated at time step $t-1$. At each time step $t$, every node $v_j \in \mathcal{S}_{t-1}$ attempts to activate inactive nodes in any hyperedge $e$ containing $v_j$, if the fraction of active nodes in $e$ is at least $\lambda$. Each eligible inactive node is then activated with probability $p$ and added to $\mathcal{S}_t$. The diffusion process terminates when no new nodes are activated in a time step, i.e., the active node set remains unchanged.

Without loss of generality, we evaluate our method under two settings of the threshold parameter $\lambda$: $\lambda = 0$ (ICRP-$0$) and $\lambda = 0.5$ (ICRP-$0.5$). This setting aims to assess the effectiveness and robustness of HIP across different diffusion mechanisms.

\begin{table*}[t]
	\centering
	\caption{Kendall's $\tau$ correlation between node influence rankings under the ICRP-0 model and those predicted by the respective method.}
	\label{tab:kend_ICRP} 
    {\Huge
    \resizebox{\linewidth}{!}{
	\begin{tabular}{cccccccccccccccc}
		\noalign{\smallskip}\hline\noalign{\smallskip}
		 &\textbf{Algorithms} & Alg & Geo & Bars & Rest & Music & iAF & iJO & SenateCm & HouseCm & HS2012 & HS2013 & ht09 & SocHam & FBTV\\
		\noalign{\smallskip}\hline\noalign{\smallskip}
		&\textbf{HIP} & \textbf{0.9136} & \textbf{0.9105} & \textbf{0.9310} & \textbf{0.9561} & \textbf{0.9532} & \textbf{0.8539} & \textbf{0.8931} & \textbf{0.9557} & \textbf{0.9150} & \textbf{0.9477} & \textbf{0.8030} & \underline{0.8182} & \textbf{0.9264} & \textbf{0.8809} \\
        
        \hline
        \multirow{4}{*}{Existing works}
		&\textbf{MPNN+LSTM} & \underline{0.8992} & \underline{0.8863} & 0.8757 & \underline{0.9211} & 0.8919 & \underline{0.7965} & \underline{0.8401} & 0.9064 & \underline{0.8887} & 0.7778 & 0.5227 & \textbf{0.8485} & \underline{0.8729} & \underline{0.8443} \\
		&\textbf{DeepIM} & 0.5173 & 0.3468 & 0.6850 & 0.5793 & 0.5762 & 0.3465 & 0.4388 & 0.1048 & 0.3567 & 0.4018 & 0.1878 & 0.1765 & 0.5528 & 0.6477 \\
		&\textbf{GBIM} & 0.2159 & 0.2063 & 0.2698 & 0.2719 & 0.2748 & 0.0453 & -0.1922 & 0.4483 & 0.2967 & 0.3856 & -0.1573 & 0.6667 & -0.0596 & 0.2012 \\
		&\textbf{ALGE} & 0.4086 & 0.3827 & 0.1576 & 0.1078 & 0.0926 & 0.1109 & 0.2619 & -0.0074 & 0.2041 & 0.2157 & 0.0000 & -0.2727 & 0.2061 & -0.0557\\
        \hline
        \multirow{2}{*}{Temporal}
		&\textbf{GLSTM} & 0.7785 & -0.5668 & 0.6097 & 0.8333 & 0.8243 & 0.7152 & 0.7060 & 0.9458 & -0.7523 & 0.8954 & 0.7614 & 0.7576 & 0.7779 & 0.3787 \\
		&\textbf{Bi-GRU} & 0.7231 & 0.8318 & 0.8716 & 0.9173 & 0.8328 & 0.7174 & 0.7951 & 0.9212 & 0.7225 & 0.7909 & 0.4924 & 0.7879 & 0.8341 & 0.7620 \\ 
        \hline
		\multirow{2}{*}{Centrality} 
        &\textbf{HCI} & 0.8180 & 0.7840 & 0.7998 & 0.6692 & 0.7844 & 0.6945 & 0.7557 & \underline{0.9360} & 0.8734 & 0.9346 & \underline{0.7614} & 0.8485 & 0.7946 & 0.6763 \\
		&\textbf{H-index} & 0.8243 & 0.8399 & \underline{0.8896} & 0.9104 & \underline{0.9247} & 0.7793 & 0.7868 & 0.8246 & 0.7866 & 0.7908 & 0.6765 & 0.7723 & 0.7571 & 0.5933 \\
		\noalign{\smallskip}\hline
	\end{tabular}
    }
    }
\end{table*}

\begin{table*}[t]
	\centering
	\caption{MSLE between node influence rankings under the ICRP-0 model and those predicted by the respective method.}
	\label{tab:MSLE} 
    {\Huge
    \resizebox{\linewidth}{!}{
	\begin{tabular}{ccccccccccccccccc}
		\noalign{\smallskip}\hline\noalign{\smallskip}
        &\textbf{Algorithm} & Alg & Geo & Bars & Rest & Music & iAF & iJO & SenateCm & HouseCm & HS2012 & HS2013 & ht09 & SocHam & FBTV\\
        \noalign{\smallskip}\hline\noalign{\smallskip}
        &\textbf{HIP} & \textbf{0.0218} & \textbf{0.0145} & \textbf{0.0132} & \textbf{0.0064} & \textbf{0.0085} & \textbf{0.0101} & \textbf{0.0087} & \textbf{0.0043} & \textbf{0.0065} & \textbf{0.0405} & \textbf{0.0018} & \textbf{0.0048} & \textbf{0.0298} & \textbf{0.0718} \\
        \hline
        \multirow{4}{*}{Existing works}
        &\textbf{MPNN+LSTM} & 0.3333 & 0.5937 & 0.4734 & 0.0878 & 0.6065 & \underline{0.0214} & \underline{0.0280} & 0.1863 & 1.5325 & 0.4893 & 0.5321 & 0.0457 & \underline{0.4930} & \underline{0.2362} \\
        &\textbf{DeepIM} & 1.8952 & 1.2130 & 0.3062 & 0.7775 & \underline{0.2084} & 0.3025 & 0.5666 & 0.6436 & 0.5439 & 0.3635 & 0.0934 & 0.2157 & 1.4306 & 0.9475 \\
        &\textbf{GBIM} & 9.0478 & 13.5505 & 8.7770 & 6.5210 & 11.7626 & 0.2076 & 0.6089 & 12.3116 & 20.2417 & 7.7285 & 16.4679 & 5.9146 & 8.3845 & 2.8244 \\
        &\textbf{ALGE} & 1.0094 & 1.2283 & 2.6166 & 0.9454 & 1.0216 & 0.2074 & 0.5504 & 0.5799 & 0.5093 & 0.7760 & 0.0362 & 0.1763 & 3.5747 & 2.7670 \\
        \hline
        \multirow{2}{*}{Temporal}
        &\textbf{GLSTM} & 2.2007 & 1.5262 & 2.6943 & 0.8960 & 1.3953 & 0.2117 & 0.6045 & 0.6174 & 1.4863 & 0.7193 & 0.3518 & 0.1645 & 3.0165 & 2.7438 \\
        &\textbf{Bi-GRU} & 1.1127 & 2.0431 & 1.2490 & 0.4080 & 1.6415 & 0.0441 & 0.0974 & 1.1982 & 4.0930 & 0.6633 & 2.1995 & 0.1271 & 2.2913 & 1.0128 \\
        \hline
        \multirow{2}{*}{Centrality}
        &\textbf{HCI} & 0.3470 & 1.1211 & \underline{0.2962} & \underline{0.0454} & 0.5317 & 0.6537 & 0.4983 & 0.1220 & 0.6250 & 0.2134 & \underline{0.0167} & \underline{0.0238} & 0.9167 & 0.4080 \\
        &\textbf{H-index} & 0.2712 & \underline{0.4982} & 0.4132 & 0.2542 & 0.1904 & 1.4098 & 1.3829 & \underline{0.0891} & \underline{0.0357} & \underline{0.0845} & 0.0965 & 0.0289 & 0.7845 & 1.1735 \\
		\noalign{\smallskip}\hline
	\end{tabular}
    }}
\end{table*}

\begin{figure}[!ht]
    \centering
    \includegraphics[width=0.85\linewidth]{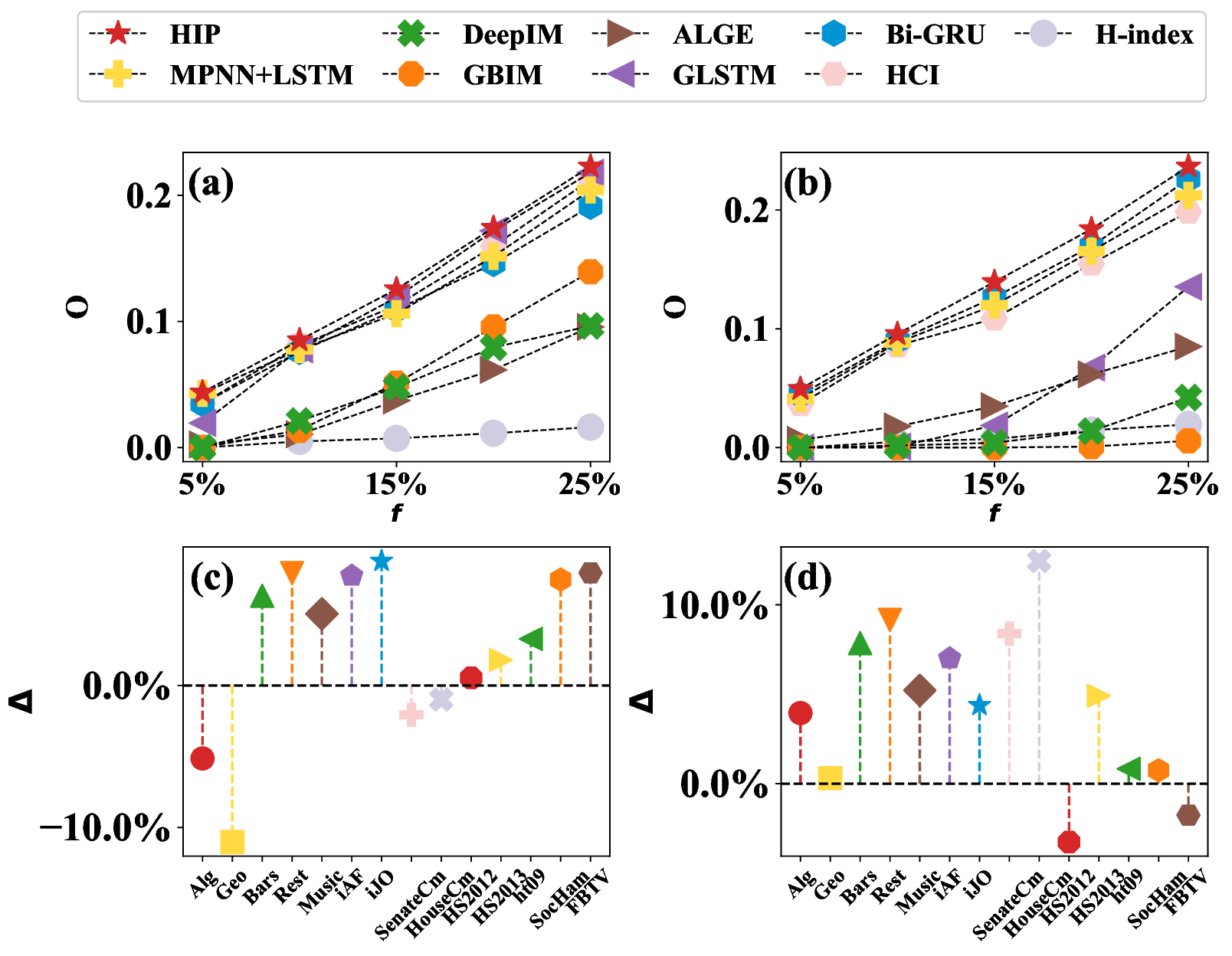}
    \caption{\textbf{(a-b)} Overlap $O$ between the top $f$ fraction of influential nodes identified by each algorithm and those ranked by the ICRP-0 and ICRP-0.5 models on the Bars-Rev dataset. \textbf{(c-d)} AUOC improvement $\Delta$ of HIP over competing methods across various hypergraphs under the ICRP-0 and ICRP-0.5 models.}
    \label{fig:Top_K}
\end{figure}

\section{Results}
\subsection{Rank Consistency}
To assess the effectiveness of our model, we report Kendall's $\tau$ of various methods under different diffusion scenarios in Tables~\ref{tab:kend_ICRP} and \ref{tab:kend_ICRP_threshold}. HIP consistently achieves the highest performance across all empirical hypergraphs, outperforming the second-best method, MPNN+LSTM, by 7\% to 11\%. DeepIM encodes nodes using one-hot vectors without considering hypergraph structure, which limits its predictive capability. GBIM learns a diffusion surrogate model based on node activation probabilities, but its reliance solely on final infection probabilities leads to poor generalization, particularly under the ICRP-$\lambda$ setting, where complex higher-order diffusion dynamics are not well captured, resulting in negative Kendall's $\tau$ values in some cases (Table~\ref{tab:kend_ICRP_threshold}). ALGE leverages conventional network-based centrality, failing to incorporate higher-order interactions inherent in hypergraphs, which leads to suboptimal performance. In contrast, GLSTM and Bi-GRU benefit from multi-scale feature encoding similar to HIP, and thus significantly outperform DeepIM, GBIM, and ALGE. Notably, Bi-GRU achieves the best results on Bars and Rest under ICRP-0.5. Heuristic methods such as HCI and H-index also maintain relatively strong performance by leveraging local degree and neighborhood structure. H-index performs particularly well on iAF, iJO, SocHam, and FBTV, which are datasets characterized by higher degree heterogeneity ($CV(\langle k \rangle)$ in Table~\ref{tab:topological properties}). This heterogeneity increases node distinguishability, and the threshold mechanism further constrains spreading ability, enhancing the importance of node degree.

In practical scenarios, top-influence nodes often play a critical role. To evaluate each method’s capability in identifying these key individuals, we assess their performance in detecting top-ranked influential nodes, as illustrated in Figure~\ref{fig:Top_K}. Figures~\ref{fig:Top_K}(a-b) present the overlap $O$ between the top-ranked nodes identified by each algorithm and those obtained using the ICRP-0 and ICRP-0.5 models on the Bars-Rev dataset. The x-axis denotes the top $f$ fraction of nodes considered $(f \in \{5\%, 10\%, 15\%, 20\%, 25\%\})$. Our results show that HIP consistently achieves superior performance, while MPNN+LSTM, Bi-GRU, and HCI demonstrate stable and comparable outcomes. Notably, although the H-index yields competitive Kendall's $\tau$ values, it performs poorly in identifying top-influence nodes. This suggests that the H-index is more adept at capturing the characteristics of nodes with moderate or low influence, rather than distinguishing the most influential ones. To further assess the robustness of our method across different hypergraphs, we compute the Area Under the Overlap Curve (AUOC)~\parencite{zhang2025locating} for each method on every dataset. We then calculate the performance margin ($\Delta$), defined as the AUOC difference between HIP and each baseline, i.e., a higher $\Delta$ indicates better performance of HIP. Figures~\ref{fig:Top_K}(c-d) illustrate the $\Delta$ values under the ICRP-0 and ICRP-0.5 models, respectively. Among all 14 hypergraphs, HIP ranks first on 10 hypergraphs under ICRP-0 and 12 under ICRP-0.5, demonstrating its strong generalization. This advantage stems from HIP's ability to capture higher-order dependencies through hypergraph neural networks, allowing for a more accurate modeling of diffusion dynamics where multi-node interactions via hyperedges are prevalent.

\subsection{Prediction Accuracy}
\begin{figure}[!ht]
\centering
	\includegraphics[width=0.8\linewidth]{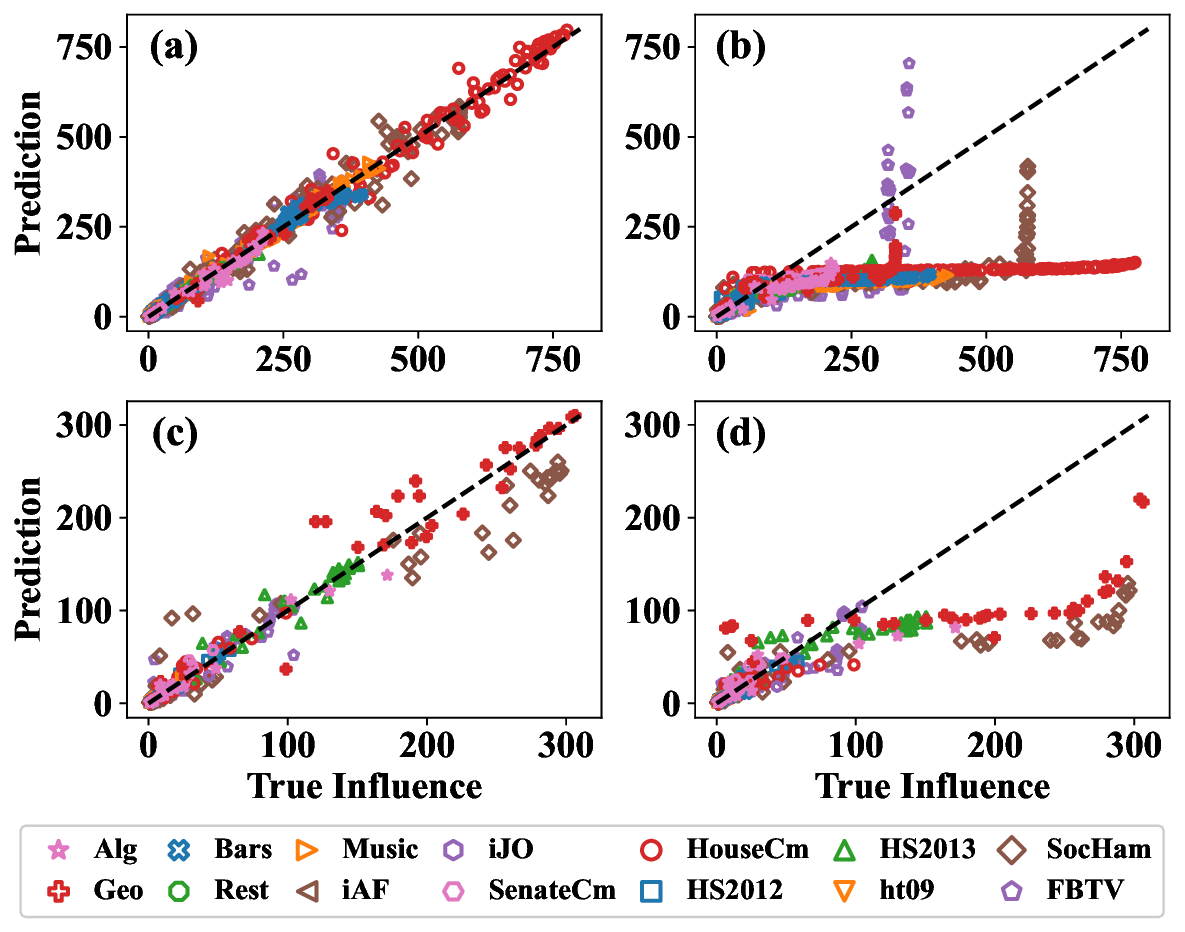}
 \caption{Comparison of predicted node influence between HIP and MPNN+LSTM under the ICRP-$\lambda$ model across empirical hypergraphs.
\textbf{(a-b)} HIP's predicted influence values versus ground-truth influence from ICRP-0 and ICRP-0.5 models.
\textbf{(c-d)} MPNN+LSTM's predicted influence values versus ground-truth influence from ICRP-0 and ICRP-0.5 models.}
 \label{fig:HGE_vs_MPNN_IC}
\end{figure}

To more accurately assess the discrepancy between predicted and actual node influence values, we compute the MSLE between the predicted influence scores from each algorithm and the ground-truth values obtained via MC simulations under the ICRP model, as reported in Table~\ref{tab:MSLE} and Table~\ref{tab:MSLE_threshold}. This numerical comparison provides a more direct evaluation than rank-based metrics. Our method, HIP, consistently achieves the lowest MSLE across all empirical hypergraphs. Notably, it reduces the MSLE by an average of 91.2\% compared to MPNN+LSTM. These quantitative results are visually supported in Figure~\ref{fig:HGE_vs_MPNN_IC}, which contrasts the predicted value of each node of HIP and MPNN+LSTM. Specifically, HIP’s predictions (Figures~\ref{fig:HGE_vs_MPNN_IC}(a) and (c)) exhibit a strong diagonal alignment with the ground-truth influence values, indicating high prediction precision. In contrast, MPNN+LSTM (Figures~\ref{fig:HGE_vs_MPNN_IC}(b) and (d)) performs reasonably well for nodes with low influence but shows increasing deviation as the predicted influence grows. Although Bi-GRU attains relatively high Kendall's $\tau$ in Table~\ref{tab:kend_ICRP_threshold}, it performs suboptimally in terms of MSLE. This suggests that while Bi-GRU effectively captures relative influence rankings, it struggles to approximate the actual magnitudes. A capacity explanation lies in the recurrent network architecture, which may suffer from vanishing gradients, impairing its capacity to model large-scale influence spread. Additionally, neural network-based approaches, such as GBIM, GLSTM, and Bi-GRU, exhibit less stable MSLE performance, with some hypergraphs yielding prediction errors surpassing those of simple centrality-based methods. This instability may stem from insufficient generalization of learned features, which hampers their regression capabilities. In contrast, centrality-based methods avoid training-induced issues such as gradient explosion and thus deliver more consistent MSLE outcomes without producing extreme predictions.

\begin{figure}[!ht]
    \centering
    \includegraphics[width=0.6\linewidth]{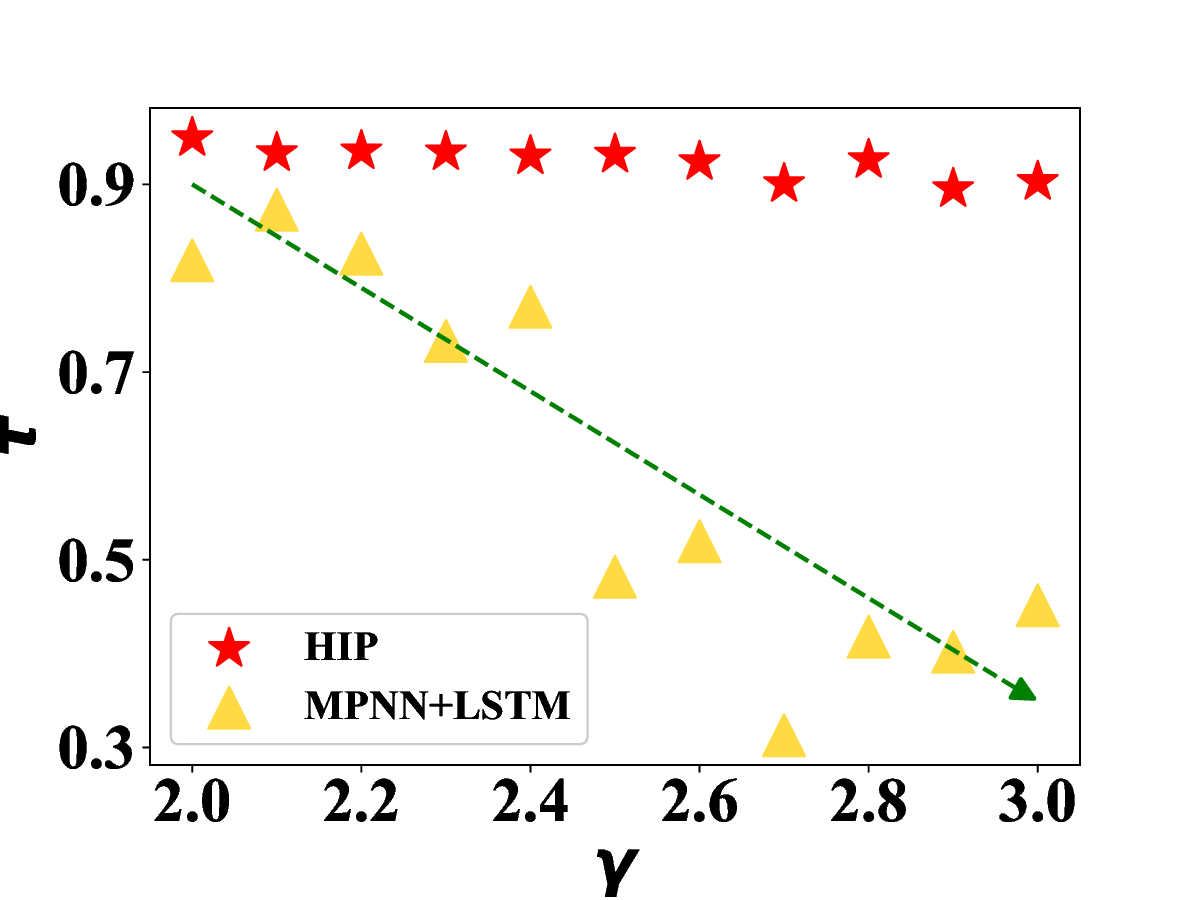}
    \caption{The robustness analysis of HIP and MPNN+LSTM across the HCL with different degree distribution exponents.}
    \label{fig:Robustness}
\end{figure}

\begin{table}[h]
	\centering
	\caption{Ablation studies of HIP.}
	\label{tab:Ablation} 

    {\fontsize{6pt}{6pt}\selectfont
    \setlength{\tabcolsep}{3.5pt} 
	\begin{tabular}{@{}ccccc@{}} 
		\hline\hline\noalign{\smallskip}	
		\multicolumn{5}{c}{\textbf{ICPR-0}} \\
		\noalign{\smallskip}\hline\noalign{\smallskip}
        \textbf{Metric} & \textbf{Kendall's $\tau$} &\textbf{Log-$R^2$} & \textbf{MSLE} & \textbf{MRLE} \\
        \noalign{\smallskip}\hline\noalign{\smallskip}
        \textbf{(w/o)Dist.} & 0.8892 (-1.66\%) & 0.9685 (-0.86\%) & 0.0249 (-43.42\%) & 0.0945 (-186.35\%) \\
        \textbf{(w/o)Cent.} & 0.8738 (-3.36\%) & 0.9423 (-3.55\%) & 0.0294 (-69.71\%) & 0.0438 (-32.79\%) \\
        \textbf{(w/o)LSTM}  & 0.8913 (-1.41\%) & 0.9650 (-1.22\%) & 0.0285 (-64.42\%) & 0.0467 (-41.62\%) \\
        \textbf{(w/o)ODE}   & 0.8801 (-2.66\%) & 0.7092 (-27.41\%) & 0.0731 (-321.60\%) & 0.1708 (-417.34\%) \\
        \textbf{HIP}  & \textbf{0.9042} & \textbf{0.9769} & \textbf{0.0173} & \textbf{0.0330} \\
        \hline\hline\noalign{\smallskip}
        \multicolumn{5}{c}{\textbf{ICPR-0.5}}\\
        \noalign{\smallskip}\hline\noalign{\smallskip}
        \textbf{(w/o)Dist.} & 0.8535 (-4.23\%) & 0.9328 (-3.66\%) & 0.0387 (-46.97\%) & 0.1075 (-135.33\%) \\
        \textbf{(w/o)Cent.} & 0.8377 (-5.99\%) & 0.9273 (-4.23\%) & 0.0475 (-80.36\%) & 0.0576 (-26.12\%) \\
        \textbf{(w/o)LSTM}  & 0.8829 (-0.92\%) & 0.9612 (-0.73\%) & 0.0342 (-29.92\%) & 0.0587 (-28.39\%) \\
        \textbf{(w/o)ODE}   & 0.8506 (-4.55\%) & 0.9211 (-4.87\%) & 0.0450 (-70.81\%) & 0.1203 (-163.17\%) \\
        \textbf{HIP}  & \textbf{0.8911} & \textbf{0.9683} & \textbf{0.0264} & \textbf{0.0457} \\
        \hline\hline\noalign{\smallskip}
	\end{tabular}
    }
\end{table}



\subsection{Ablation Study}
To assess the contribution of each component in HIP, we perform a comprehensive ablation study by individually removing the distance features, centrality features, LSTM, and Neural ODE modules. Model performance is evaluated using four metrics: Kendall's $\tau$, Log-$R^2$, MSLE, and MRLE. As shown in Table~\ref{tab:Ablation}, we report both the absolute values of these metrics and their relative degradation (in parentheses) compared to the full HIP model under the ICRP model. All results represent averages across 14 empirical hypergraphs. Among the components, the Neural ODE module proves to be the most critical, with its removal causing the largest performance drop, particularly under the ICRP-0 model, where MSLE and MRLE increase by 321.6\% and 417.34\%, respectively. In contrast, removing the LSTM module leads to minimal performance changes. Centrality and distance features exhibit comparable effects on Kendall's $\tau$ and Log-$R^2$. Notably, the differing behaviors of MSLE and MRLE reveal complementary aspects of feature contribution: MSLE, which penalizes squared logarithmic errors, is more sensitive to small-value inaccuracies and shows greater degradation when centrality features are excluded; MRLE, which emphasizes relative errors, is more influenced by large-value discrepancies and thus drops more significantly when distance features are removed. These findings underscore the distinct yet complementary roles of centrality and distance features in capturing the full distribution of node influence.

\subsection{Robustness Analysis via Synthetic Hypergraphs}
To assess the robustness of our method to degree heterogeneity, we compare the Kendall's $\tau$ of HIP and the strongest baseline, MPNN+LSTM, on synthetic hypergraphs generated by the HyperCL model with systematically tuned degree distribution exponents. To generate a synthetic hypergraph $H(V, E)$ with $N$ nodes and $M$ hyperedges, we first sample node hyperdegrees ${k^H_1, \cdots, k^H_N}$ from a power-law distribution $p(k^H) \sim (k^H)^{-\gamma}$, where $\gamma$ controls the degree heterogeneity. Each hyperedge size ${k^E_1, \cdots, k^E_M}$ is drawn uniformly from ${2, 3, 4, 5}$. For each hyperedge $e_i$, nodes are added without duplication according to probability $\frac{k^H(v_j)}{\sum_{j=1}^{N}k^H(v_j)}$ until its size reaches $k^E_i$. The process continues until all $M$ hyperedges are constructed.

As $\gamma$ increases, the hypergraph becomes more degree-homogeneous, which may pose additional challenges for distinguishing node influence. We vary $\gamma$ from 2.0 to 3.0 in increments of 0.1. As illustrated in Figure~\ref{fig:Robustness}, HIP maintains stable performance across the entire range of $\gamma$, while the Kendall's $\tau$ of MPNN+LSTM declines significantly. When $\gamma \geq 2.5$, MPNN+LSTM nearly fails to capture the correct influence ranking. These results demonstrate that HIP exhibits greater robustness and scalability across hypergraphs with varying structural heterogeneity.



\begin{table}[t]
	\centering
	\caption{Compatibility of HIP model.}
	\label{tab:module updates} 

    {\fontsize{6pt}{6pt}\selectfont
    \setlength{\tabcolsep}{1.5pt} 
	\begin{tabular}{@{}c|cccc|cccc@{}} 
		\hline\hline\noalign{\smallskip}	
		& \multicolumn{4}{|c|}{\textbf{ICPR-0}} & \multicolumn{4}{c}{\textbf{ICPR-0.5}} \\
		\noalign{\smallskip}\hline\noalign{\smallskip}
        \textbf{Metric} & \textbf{Kendall's $\tau$} &\textbf{Log-$R^2$} & \textbf{MSLE} & \textbf{MRLE} & \textbf{Kendall's $\tau$} &\textbf{Log-$R^2$} & \textbf{MSLE} & \textbf{MRLE}\\
        \noalign{\smallskip}\hline\noalign{\smallskip}
        HIP  & \textbf{0.9042} & 0.9769 & 0.0173 & 0.0330 & 0.8911 & 0.9683 & 0.0264 & 0.0457 \\
        HIP (xLSTM)  & 0.8954 & 0.9631 & 0.0245 & 0.0380 & 0.8877 & 0.9615 & 0.0316 & 0.0536 \\
        HIP (DPHGNN)  & 0.9023 & \textbf{0.9791} & \textbf{0.0145} & \textbf{0.0308} & \textbf{0.9034} & \textbf{0.9766} & \textbf{0.0176} & \textbf{0.0416} \\
        HIP (DP+xL)  & 0.8918 & 0.9680 & 0.0174 & 0.0339 & 0.8914 & 0.9656 & 0.0250 &0.0471\\
        \hline\hline\noalign{\smallskip}
	\end{tabular}
    }
\end{table}

\subsection{Module Compatibility Analysis}
To assess the generalizability of the HIP framework, we substitute its core components with more advanced modules, i.e., replace HGNN and LSTM modules with Dual Perspective Hypergraph Neural Networks (DPHGNN)~\parencite{saxena2024dphgnn} and Extended Long Short-Term Memory (xLSTM)~\parencite{beck2024xlstm}. The corresponding results are as summarized in Table~\ref{tab:module updates}. Incorporating DPHGNN, referred to as HIP-DPHGNN, enhances the model's ability to capture complex hypergraph node representations, resulting in a slight improvement in prediction performance. Conversely, replacing LSTM with xLSTM, denoted as HIP-xLSTM, introduces a marginal performance drop; nevertheless, HIP-xLSTM still achieves competitive results, underscoring the robustness and modularity of the HIP architecture across different neural designs.

\section{Conclusion and Future work}
Node influence prediction is central to spreading dynamics, but conventional networks often overlook higher-order interactions captured by hypergraphs, which also pose new challenges. To this end, we propose HIP, an end-to-end general framework that integrates structural feature encoding with temporal modeling to accurately estimate node influence. HIP leverages a range of centrality measures and a temporally reinterpreted distance matrix to capture multi-scale structural information. The conventional neural network module aggregates node features, while recurrent neural network and Neural ODE modules enhance the model's ability to capture temporal dependencies. Extensive experiments on empirical hypergraphs demonstrate that the HIP framework consistently outperforms state-of-the-art methods in ranking accuracy, prediction fidelity, and identification of top influential nodes.

Looking ahead, more advanced message passing or recurrent neural network architectures can be seamlessly incorporated into HIP, owing to its high compatibility and modular design. Moreover, the integration of large language models into influence modeling offers a promising new direction.



\clearpage
\printbibliography

\clearpage
\appendix
\renewcommand{\thetable}{A.\arabic{table}}
\setcounter{table}{0}

\section{Appendix}

\subsection{Hypergraph Centrality Measures}
\label{appendix:centrality}
\begin{itemize}
\item \textbf{Degree Centrality (DC)} quantifies the node influence by the number of neighbors, i.e., $k_i = \sum^N_{j=1}\mathbf{A}_{ij}$.
\item \textbf{Hyperdegree Centrality (HDC)} assumes that nodes participating in more hyperedges are more likely to accelerate information diffusion. The HDC of node $v_i$ is $k^H_i = \sum^M_{j=1}\mathbf{I}_{ij}$.
\item \textbf{Vector Centrality (VC)}~\parencite{kovalenko2022vector} evaluates node influence by incorporating eigenvector centrality of hyperedges. Specifically, the hypergraph $H$ is transformed into its line graph $L(H)$, where each hyperedge in $H$ corresponds to a node in $L(H)$. Eigenvector centrality is then computed on $L(H)$, and the resulting centrality score of each hyperedge is uniformly allocated to its constituent nodes. 
Accordingly, the VC score of node $v_i$ is given by$c_i=\sum_{k=1}^{K}c_{ik}$, where $c_{ik}$ denotes the share of centrality assigned from hyperedge $e_{ik} \in E_i$.
\item \textbf{Hypergraph Gravity Centrality (HGC)}~\parencite{xie2023vital} quantifies node influence based on a gravitational analogy, considering both node degree (mass) and pairwise proximity (distance). In this formulation, nodes with higher degrees and shorter distances to others exert a stronger influence. The HGC of a node $v_i$ is defined as: $g_i = \sum_{i \neq j}\frac{k_ik_j}{(d_v(i,j))^2}$.
\item \textbf{Harmonic Closeness Centrality (HCC)}~\parencite{aksoy2020hypernetwork} assumes that a hyperedge is more critical if its average distance to other hyperedges is smaller. The HCC of a hyperedge $e_p$ is defined as: $h^e_p=\frac{1}{M-1}\sum_{e_p,e_q\in E, p\neq q}\frac{1}{d_e(p,q)}$, where $d_e(p, q)$ denotes the distance between hyperedges $e_p$ and $e_q$.
To derive node-level centrality, each hyperedge's HCC score is evenly distributed among its member nodes. Therefore, the HCC value of a node $v_i$ is defined as: $h_i = \sum_{e_p \in E_i}\frac{h^e_p}{|e_p|}$, where $E_i$ is the set of hyperedges containing $v_i$.
\end{itemize}

\begin{table*}[h]
	\centering
	\caption{Kendall's $\tau$ correlation between node influence rankings under the ICRP-0.5 model and those predicted by the respective method.}
	\label{tab:kend_ICRP_threshold} 
    {\Huge
    \resizebox{\linewidth}{!}{
	\begin{tabular}{cccccccccccccccccc}
		\noalign{\smallskip}\hline\noalign{\smallskip}
		&\textbf{Algorithms} & Alg & Geo & Bars & Rest & Music & iAF & iJO & SenateCm & HouseCm & HS2012 & HS2013 & ht09 & SocHam & FBTV\\
		\noalign{\smallskip}\hline\noalign{\smallskip}
		&\textbf{HIP}        & \textbf{0.9092} & \textbf{0.9153} & \underline{0.9414} & \textbf{0.9420} & \textbf{0.9432} & \textbf{0.8474} & \underline{0.8199} & \textbf{0.9606} & \textbf{0.9118} & \textbf{0.9346} & \textbf{0.8598} & \textbf{0.7879} & \textbf{0.8315} & \textbf{0.8713} \\
        \hline
        \multirow{4}{*}{Existing works}
		&\textbf{MPNN+LSTM} & \underline{0.9048} & \underline{0.9141} & 0.8735 & 0.8982 & 0.9023 & 0.4590 & 0.5882 & 0.9015 & \underline{0.8514} & \underline{0.9216} & 0.7311 & \underline{0.7273} & 0.7704 & 0.7958 \\
		&\textbf{DeepIM}    & 0.5591 & 0.3729 & 0.5617 & 0.5269 & 0.4469 & 0.2564 & 0.3773 & 0.3579 & 0.2429 & 0.4997 & 0.3721 & -0.0053 & 0.4148 & 0.3633 \\
		&\textbf{GBIM}      & 0.7099 & 0.7649 & -0.5929 & -0.6399 & 0.5661 & -0.1873 & -0.3332 & 0.7931 & 0.6942 & 0.7386 & 0.1708 & 0.5758 & -0.2312 & -0.4388 \\
		&\textbf{ALGE}      & 0.4529 & 0.4541 & 0.2100 & 0.0859 & 0.4129 & 0.0606 & 0.1340 & -0.0173 & 0.2083 & 0.1895 & 0.0076 & 0.5758 & 0.2438 & -0.0375 \\
        \hline
        \multirow{2}{*}{Temporal}
		&\textbf{GLSTM}     & 0.4197 & -0.0974 & 0.2939 & -0.2777 & 0.2526 & 0.2367 & -0.1722 & 0.8473 & 0.3302 & 0.7255 & 0.5985 & 0.5455 & 0.7094 & 0.5286 \\
		&\textbf{Bi-GRU}    & 0.8804 & 0.8306 & \textbf{0.9420} & \textbf{0.9420} & \underline{0.9387} & 0.5075 & 0.6444 & \underline{0.9458} & 0.7867 & 0.9085 & 0.7273 & \underline{0.7273} & 0.7483 & 0.8020 \\
        \hline
        \multirow{2}{*}{Centrality}
		&\textbf{HCI}        & 0.7911 & 0.7617 & 0.8352 & 0.7013 & 0.7578 & 0.7013 & 0.7392 & 0.9261 & 0.7864 & 0.9085 & \underline{0.7689} & \underline{0.7273} & 0.7585 & 0.7557 \\
		&\textbf{H-index}   & 0.8683 & 0.8574 & 0.9246 & 0.9103 & 0.8979 & \underline{0.8395} & \textbf{0.8428} & 0.8014 & 0.6981 & 0.7509 & 0.6577 & 0.7263 & \underline{0.7764} & \underline{0.8584} \\
		\noalign{\smallskip}\hline
	\end{tabular}
    }}
\end{table*}

\begin{table*}[t]
	\centering
	\caption{MSLE between node influence rankings under the ICRP-0.5 model and those predicted by the respective method.}
	\label{tab:MSLE_threshold} 
    {\Huge
    \resizebox{\linewidth}{!}{
	\begin{tabular}{ccccccccccccccccc}
		\noalign{\smallskip}\hline\noalign{\smallskip}
         &\textbf{Algorithm} & Alg & Geo & Bars & Rest & Music & iAF & iJO & SenateCm & HouseCm & HS2012 & HS2013 & ht09 & SocHam & FBTV\\
        \noalign{\smallskip}\hline\noalign{\smallskip}
        &\textbf{HIP}         & \textbf{0.0414} & \textbf{0.0543} & \textbf{0.0017} & \textbf{0.0014} & \textbf{0.0032} & \textbf{0.0002} & \textbf{0.0021} & \textbf{0.0228} & \textbf{0.0338} & \textbf{0.0337} & \textbf{0.0258} & \textbf{0.0341} & \textbf{0.0693} & \underline{0.0452} \\
        \hline
        \multirow{4}{*}{Existing works}
        &\textbf{MPNN+LSTM}  & \underline{0.0835} & \underline{0.4893} & \underline{0.0090} & \underline{0.0067} & \underline{0.0085} & 0.0007 & \underline{0.0088} & 0.0572 & \underline{0.0606} & \underline{0.0885} & 0.1931 & 0.0885 & \underline{0.1484} & \textbf{0.0394} \\
        &\textbf{DeepIM}      & 1.9580 & 4.1137 & 0.1771 & 0.1021 & 0.1662 & 0.0382 & 0.1596 & 1.1048 & 0.8721 & 0.7792 & 0.7199 & 0.9481 & 1.8955 & 0.9036 \\
        &\textbf{GBIM}        & 2.9612 & 10.3429 & 0.7920 & 0.2337 & 0.8523 & 0.0247 & 0.7585 & 0.7752 & 4.2661 & 1.2606 & 3.0271 & 0.5909 & 15.0049 & 1.0982 \\
        &\textbf{ALGE}        & 1.0013 & 2.4588 & 0.1874 & 0.1343 & 0.1355 & 0.0052 & 0.1357 & 0.8271 & 0.7190 & 1.2979 & 0.6378 & 0.5036 & 1.6432 & 1.1046 \\
        \hline
        \multirow{2}{*}{Temporal}
        &\textbf{GLSTM}       & 1.4349 & 4.6455 & 0.1850 & 0.1270 & 0.1835 & 0.0047 & 0.1348 & 0.8662 & 0.7776 & 1.1527 & 0.6255 & 0.4275 & 1.8517 & 1.1069 \\
        &\textbf{Bi-GRU}      & 0.2849 & 1.6541 & 0.0154 & 0.0150 & 0.0224 & \underline{0.0004} & 0.0256 & \underline{0.0358} & 0.1766 & 0.1422 & 0.6749 & \underline{0.0694} & 0.5685 & 0.2598 \\
        \hline
        \multirow{2}{*}{Centrality}
        &\textbf{HCI}          & 0.7966 & 0.5381 & 1.4857 & 1.2288 & 1.1781 & 1.6859 & 1.6366 & 0.6493 & 0.7946 & 0.3040 & \underline{0.0394} & 0.4892 & 0.6740 & 0.5507 \\
        &\textbf{H-index}     & 0.9109 & 0.9962 & 1.5119 & 1.2167 & 1.3996 & 2.0122 & 2.0370 & 0.6702 & 1.0329 & 0.2089 & 0.1752 & 0.1148 & 1.8053 & 1.8935 \\
		\noalign{\smallskip}\hline
	\end{tabular}
    }}
\end{table*}

\end{document}